\documentstyle[11pt,paspconf,epsf]{article}

\begin{document}

\title{The PMN Selected Survey for Southern Gravitational Lenses}

\author{O.~R.~Prouton \& S.~J.~Warren (Imperial College, London, U.K.), \\ A.~E.~Wright (ATNF, Parkes, Australia)}

\begin{abstract}

We are surveying the region $-87^\circ < \delta < -38^{\circ}.5$ for multiple imaging at scales $1'' \leq \Delta \theta \leq 15''$. Candidate gravitational lenses were selected from an existing dataset of Australia Telescope Compact Array measurements of 2178 flat (4.80/8.64~GHz) spectrum PMN sources. Selection criteria were optimised using tests on simulated lens datasets.

\end{abstract}



The Parkes-MIT-NRAO (PMN) 4.85~GHz survey (Griffith~\&~Wright~1993) covered the entire southern sky to a limiting flux density of 30~mJy (varying slightly with declination); it increased the number of known southern radio sources by a factor of approximately 6. Wright et al. (1996) made simultaneous 4.80/8.64~GHz Australia Telescope Compact Array (ATCA) measurements of a complete sample of 8068 PMN sources defined by the flux density limits $S_{4.85~\rm{GHz}} \geq 70$~mJy $(-73^\circ < \delta < -38^\circ.5)$, $S_{4.85~\rm{GHz}} \geq 50$~mJy $(-87^\circ < \delta < -73^\circ)$ and the requirement that galactic latitude $|b| \geq 2^\circ$. The Wright et al. ATCA observations resulted in: positions accurate to (standard error) $0.6''$; spectral data and structural information of resolution approximately $1''$. This dataset is the largest deep, complete, unbiased, dual-frequency, interferometric catalogue of southern radio sources. It covers $19~\%$ of the sky, yet this area contains only 1 of the 39 previously known cases of gravitational lensing classed as \emph{probable\ lenses} by the \emph{CASTLES}\footnote{$\rm{http://cfa-www.harvard.edu/castles/castles.edu}$} collaboration (Figure 1). The resolution ($\sim 1''$) samples the peak of the image separation histogram for the 14 JVAS/CLASS lenses published prior to these proceedings (Browne~et~al.,~1998, Koopmans~et~al.,~1999, Myers~et~al.,~1999). The 2178 flat--spectrum sources in these data constitute our finding survey.


Candidate gravitational lenses were selected using an automated routine, based on objective criteria, designed to reduce sample contamination by radio galaxies and objects within, or behind, the Galactic plane. Radio galaxies are often observed to exhibit steepening radio spectral indices with increasing separation from the core (e.g. Wiita \& Gopal-Krishna, 1990), objects showing such spectral steepening were not included in the sample.  Simulated ATCA observations of known northern hemisphere lenses (transposed to declinations within the surveyed region) were used to test and refine candidate selection.

The 110 flat-spectrum objects satisfying our selection criteria of component fluxes $S_{\nu}> 5 \sigma~(\sim~30$~mJy) (at both 4.80 and 8.64~GHz) and $|b| \geq 5^\circ$ were chosen for further study at radio, infrared and optical wavelengths. New simultaneous 4.80/8.64 GHz, dual linear polarisation ATCA observations of each lens candidate have been made. These new data (of resolution $\sim 1''$ at 8.64~GHz) exhibit a factor 8 improvement in S:N, and vastly improved coverage of the interferometric u-v plane, over data in hand. Total intensity and polarisation maps of all objects have been made and examined. Re-running the automated lens candidate finder rejected half of the observed objects. We have obtained deep near infrared $K_N$-Band images of 35 sources remaining in the lens candidate sample, deep optical B-Band images (and colours) for 19 of these objects and 1 spectrum.

\begin{figure}
\caption{\small{Aitoff projection showing the positions of known gravitational lenses, their discovery method and the sky coverage of the current major surveys for flat radio spectrum gravitational lenses. The hatched region is covered by this survey (the region shaded black represents the un-surveyed area of radius $3^\circ$ centered on the southern celestial pole). The region bounded by solid black lines is the area ($0^\circ > \delta > -40^\circ$, $|b| > 10^\circ$) surveyed by the MIT group  (Winn et al., these proceedings), the regions of the northern sky for which $|b| > 10^\circ$ (dashed line) are covered by CLASS (Browne et al., these proceedings).}}
\plotfiddle{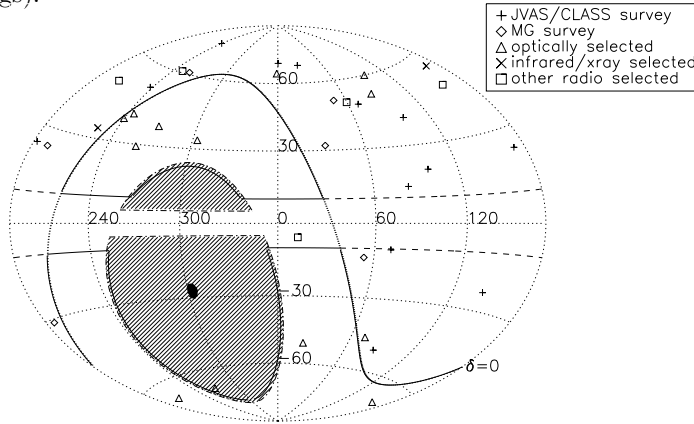}{5.15cm}{90}{40}{40}{125}{-50}
\end{figure}



\end{document}